\documentclass[12pt]{article}
\usepackage{amsfonts} 
\usepackage{a4}
\usepackage{epsfig}
\usepackage{latexsym}
\begin{document}

\title {Some remarks on black hole temperature and the second law of thermodynamics}
\author{Marco Scandurra \thanks{e-mail: mscandurra@bert.uchicago.edu; scandurr@df.unipi.it}\\
{\itshape University of Pisa}\\
{\itshape Dipartimento di Fisica}\\
{\itshape Via Buonarroti 2, I-56127 Pisa, Italy}}\maketitle

\begin{abstract}
I present a  formulation of the second law of thermodynamics in the presence of black holes which makes use of the efficiency of an ideal machine extracting heat cyclically from a black hole. 
The Carnot coefficient is found and it is shown to be a simple function of the mass.
\end{abstract}

Black holes are known to radiate energy thermally  with a temperature given by a simple function of the mass of the black hole\cite{Hawking}.  
In the case of an uncharged, non rotating black hole the temperature is given by
\begin{equation}
T(M)=\frac{\hbar c^3}{8\pi k_B G M}\ .
\end{equation} 

The radiation has a quantum origin: it arises from the distortion of the vacuum fluctuations of the quantum fields near the event horizon \cite{Hawking,Mostepanenko2}.
Classical black holes are found to obey ordinary laws of thermodynamics \cite{Hawking2}. 
Furthermore, there are  reasons to believe that radiating  black holes obey a generalized second law of thermodynamics, 
stating that the sum of the entropy of the black hole $\frac 14 A$ and  the ordinary entropy outside it $S$ never decreases \cite{Bekenstein}. 
The role of the entropy inside a black hole is played by the event horizon area $A$.

Since a black hole is fully comparable to a black body emitting real radiation, one could imagine that a body at a lower temperature located in its vicinity could absorb part of this radiation. 
However, if our goal is to extract energy from the black hole, we would find that only a limited quantity of energy available for doing work could be obtained by this method. 
The impossibility of  extracting energy from a single heat bath is a strict consequence of the second law of thermodynamics.

In this note I would like to show how the second law of thermodynamics
in the context of black hole physics can be formulated\footnote{It is in fact only a different formulation, not a proof of its validity} 
by means of the Carnot coefficient $\eta$, which is a well known parameter in classical thermodynamics  expressing the efficiency of a thermal machine exploiting two heat reservoirs.

In the formulation by Thomson, the second law of thermodynamics states that a machine can not extract energy cyclically from a heat reservoir without changing anything in the external environment. 
At least two heat reservoirs at temperatures $T_1$ and $T_2$ are needed in order to transform heat into work. 
Then a machine can mine energy cyclically from the hotter reservoir $T_1$ by releasing a part of it
to the cooler one $T_2$. The maximum possible efficiency of a cycle is given by the Carnot coefficient
\begin{equation}
\eta =\frac{T_1-T_2}{T_1}\ .
\end{equation}

If Hawking radiation has to be considered as real and  if  the generalized second law of thermodynamics holds, 
then a coefficient $\eta_{BH}$ should exist and it should express the efficiency of an ideal  machine which  extracts energy cyclically from a black hole. 
More exactly, we imagine a machine which exploits two black holes with masses $M_1$ and $M_2$. 
The machine takes heat from the black hole with smaller mass $M_1$ and releases a part of it to the black hole with larger mass $M_2$. 
Only a fraction of the extracted energy can be transformed into work. 
Using the expression for the temperature given in (1) for masses  $M_1$ and $M_2$ and inserting it into formula (2)  we obtain
\begin{equation}
\eta_{BH} =\frac{M_2-M_1}{M_2}\ <\ 1\ .
\end{equation}
This has to be considered as the Carnot coefficient in black hole mechanics. 
The fact that $\eta_{BH}$ is always smaller than one is equivalent to asserting  that the entropy of the system can never decrease. 
In fact,  if we could extract energy from a single black hole we would decrease its entropy (though we would increase its temperature), without causing any other modification in the vicinity.

\begin{Large}
\vspace{0.8cm}
\noindent {\bf Acknowledgment} \end{Large}

\vspace{0.4cm}
\noindent I would like to thank Prof. Wald and all the members of the Relativity Group for their hospitality at the Enrico Fermi Institute.

\end{document}